\documentclass[mathleft]{an}
\usepackage{graphicx}
\usepackage{times}
\begin{document}

\Pagespan{789}{}
\Yearpublication{2007}%
\Yearsubmission{2007}%
\Month{12}%
\Volume{328}%
\Issue{10}%
\DOI{10.1002/asna.200710887}%

\title{3-D non-linear evolution of a magnetic flux tube in a spherical shell: the isentropic case}

\author{L. Jouve \fnmsep\thanks{Corresponding author:
  \email{ljouve@cea.fr}\newline}
\and  A.S. Brun
}
\titlerunning{Rising flux tubes}
\authorrunning{L. Jouve \& A. S.Brun}
\institute{Laboratoire AIM, CEA/DSM-CNRS-Universit\'e Paris Diderot, DAPNIA/SAp, 91191 Gif sur Yvette, France
}

\received{30 May 2005}
\accepted{11 Nov 2005}
\publonline{later}


\abstract{We present recent 3-D MHD numerical simulations of the non-linear dynamical
evolution of magnetic flux tubes in an adiabatically stratified convection zone in spherical geometry, using the anelastic spherical
harmonic (ASH) code. We seek to understand the mechanism of emergence
of strong toroidal fields from the base of 
the solar convection zone to the solar surface as active regions. We confirm the results obtained in cartesian geometry that flux tubes that are not twisted split into two counter vortices before reaching the top of the convection zone. Moreover, we find that twisted tubes undergo the poleward-slip instability due to an unbalanced magnetic curvature force which gives the tube a poleward motion both in the non-rotating and in the rotating case. This poleward drift is found to be more pronounced on tubes originally located at high latitudes. Finally, rotation is found to decrease the rise velocity of the flux tubes through the convection zone, especially when the tube is introduced at low latitudes.
}

\maketitle

\section{Introduction}

The Sun possesses a large variety of magnetic phenomena such as sunspots following an 11-yr cycle, explosive flares or CME's. Such active regions on the solar surface are believed to take their origin from strong toroidal fields created at the base of the convection zone (CZ). We thus need to understand the rising mechanisms of strong toroidal structures through the CZ. Many models carried out since the 80's relied on the assumption that toroidal flux is organized in the form of discrete flux tubes which will rise cohesively from the base of the CZ up to the solar surface (see Cattaneo, Brummell \& Cline  2006 however for a less idealized idea of the topology of buoyant flux structures). The first models using "thin flux tube approximation" (Spruit 1981) showed that the initial strength of magnetic field was an important parameter in the evolution of the tube. They also showed that an uncompensated magnetic curvature force could deflect the trajectory of the tube poleward, making the tube emerge at very high latitudes, contrary to what is observed in the Sun (Spruit \& van Ballegooijen 1982; Moreno-Insertis, Sch{\"u}ssler \& Ferriz Mas 1992). These results were also obtained in axisymmetric simulations of flux tubes in a rotating background (Choudhuri \& Gilman 
1987). 
More sophisticated multidimensional models in cartesian geometry were then developed (see review of Fan, 2004). These calculations showed that  a sufficient twist is needed to maintain the coherence of the tube during its rise (e.g. Emonet \& Moreno-Insertis 1998).    
 We here present the first attempt to study the 3-D full MHD evolution of toroidal flux tubes in spherical geometry, focusing on the effects of the twist of the field lines, of rotation and of the initial latitude of the tube on its emergence.

 The paper is organized in the following manner. In section 2, we present the initialization of the model; in section 3 we discuss the effects of the twist of the field lines, of a background rotation and of the initial latitude on the dynamical evolution of the flux tube and we conclude in section 4.

\section{The model: initial conditions and background state}

We use the ASH code (anelastic spherical harmonic, see Brun, Miesch  
\& Toomre 2004 for details) to solve in three dimensions the MHD equations
 in a convective spherical shell. ASH uses the anelastic approximation that filters sou-
 nd waves without suppressing the effects of density stratification. Solar values are used for the rotation rate and luminosity, and the initial stratification is obtained from a one-dimensional solar structure model.
All the simulations are computed 
in a shell with $r\in[0.7,0.96]R_ {\odot}$ with the resolution $N_r=N_\theta=N_{\phi}/2=256$.

To compute the model, we introduce at the  
starting time a torus of magnetic field in entropy equilibrium with the surrounding medium at the  
base of the computational domain and we let this MHD-simulation  
evolve.  In this first paper, we will not address how such coherent idealized magnetic flux tubes are created within the Sun. This regular axisymmetric magnetic structure is embedded in an unmagnetized stratified medium. We choose for simplicity to keep the tube axisymmetric. We could have introduced an inhomogenous tube with respect to longitude in our 3-D simulation, to study for example $\Omega$-loops emergence. However, our main goal here is to have a reference case to which we will compare our fully convective non isentropic simulations in a follow up paper (Jouve \& Brun 2008).
 In order to keep a divergenceless magnetic field, we use a toroidal-poloidal decomposition,

\begin{equation}
{\bf B}={\bf \nabla\times\nabla\times}(C {\bf e}_r) +{\bf \nabla\times}(A {\bf e}_r) 
\end{equation}

\noindent the expressions used for the potentials $A$ and $C$ for the flux tubes are: 

\begin{eqnarray}
A=-B_{0} &r& \exp\left[-\left(\frac{r-R_t}{a}\right)^2\right]  \\ \nonumber
& & \,\,\,\,\, \,\,\, \,\,\,  \times \left[1+\tanh\left(2\frac{\theta-\theta_t}{a/R_t}\right)\right]
\end{eqnarray}

\begin{eqnarray}
C=-B_{0} \frac{a^2}{2}& q& \exp\left[-\left(\frac{r-R_t}{a}\right)^2\right]  \\ \nonumber
& &  \,\,\,\,\, \,\,\, \,\,\,  \times \left[1+\tanh\left(2\frac{\theta-\theta_t}{a/R_t}\right)\right]
\end{eqnarray}

\noindent where $B_0$ is a measure of the initial field strength, $a$ is the tube radius, $(R_t,\theta_t)$ is the position of the tube center and $q$ is the twist parameter.

The entropy gradient of the background state is set to zero so that we do 
not trigger the convection instability in this study. The density contrast between the top and the bottom of the CZ is here about 40 and the tube radius is set to $2\times10^9 \, \rm cm$, about 0.4 times the pressure scale height at the base of the CZ.
We compute an  
untwisted case (the initial field is exclusively oriented in the  
direction of the tube, i.e. $q$=0), a twisted case (with a twist above the threshold of Eq. \ref{eqth}), cases with tubes located at three different latitudes (namely at 15${\degr}$, 45${\degr}$ and 60${\degr}$) in a rigidly rotating shell and we will compare them with a non-rotating computation.

\section{Dynamical evolution of the tube}

Since the tube is introduced in entropy and pressure equilibria it possesses a lower density than the surrounding me-
dium. Consequently, the tube is buoyant and begins to rise.

\subsection{Untwisted vs twisted tubes}
\begin{figure}[h]
    \centering
  \includegraphics[width=9.5cm]{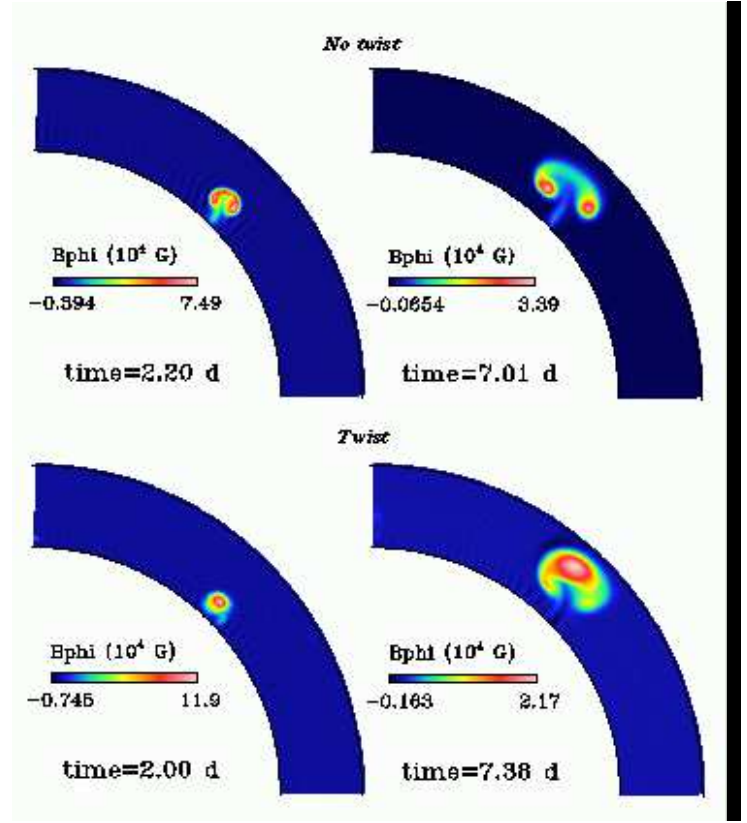}
      \caption{The four panels are snapshots of $B_{\phi}$ 
       for the untwisted case (two upper panels) and  
the twisted case (two lower panels), cut at a specific longitude, shown in a northern quadrant of the sphere, after 2 days (left panels) and 7 days (right panels) of evolution.}
        \label{figure_twist}
    \end{figure}

Figure $\ref{figure_twist}$ shows the results of the evolution, in a shell rotating at the rate $\Omega_{0}=2.6\times10^{-6}\,\rm rad.s^{-1}$, of a  
flux tube with an initial intensity of $1.8\times10^5 \rm G$ and located at a latitude of 45$\degr$ in 
both the untwisted case and in the twisted case (with a twist above the threshold, see discussion below), at two instants of the dynamical evolution.

We clearly see in the case without twist (top row of Fig. $\ref 
{figure_twist}$), the formation of two  
counter vortices, splitting the tube in two parts while it rises. As  
shown in Emonet \& Moreno-Insertis (1998), this splitting can be  
understood by studying the equation for azimuthal vorticity.
If we do not have any twist, the projection of the Lorentz force on the  
transverse plane, which constitutes a sink of vorticity, vanishes. 
Consequently, no processes can counteract the  
generation of vorticity due to the gravitational torque which results in 
the formation of two counter vortices as the tube rises.
In the twisted case (lower row of Fig. $\ref 
{figure_twist}$), with a twist above a certain threshold, magnetic  
tension prevents vorticity to be created in the main body of the  
tube by counteracting the deformation of the field lines: the tube remains coherent during its rise. 

 By setting the gravitational torque to be equal to the projection of the Lorentz force in the equation for the azimuthal vorticity, we can determine the threshold above which the twist of the field lines can counteract the creation of two counter vortices inside the tube. Defining the pitch angle $\psi$ as the angle between the direction of the vector magnetic field and the longitudinal direction, we derive:
\begin{equation}
\label{eqth}
\sin\psi=\frac{\sqrt{(B_r^2+B_\theta^2)}}{B} \geq \sqrt{\frac{a}{H_p}}\times\sqrt{\left|\frac{\Delta\rho}{\rho}\right| \frac{\beta}{2}}
\end{equation}
where $H_p$ is the pressure scale height at the base of the CZ, $\Delta\rho/\rho$ is the density deficit inside the tube compared to the background stratification
divided by the background density at the tube center and $\beta$ is the plasma-$\beta$ associated with the tube. In our case, the threshold value is equal to $0.32$ (corresponding to a pitch angle of 18.7$\degr$). In the twisted case, we then use for $\sin\psi$ a value of $0.52$ (corresponding to a pitch angle of 31.3$\degr$), i.e. well above the threshold, so that the tube is able to rise cohesively through the entire CZ. 

This amount of twist can be related to the winding degree $n$ of the field lines by the following formula:
\begin{equation}
n=\frac{\pi R_t \sin\theta_t}{2 a}\tan\psi
\end{equation}
which gives in our case a winding degree of about 17. At the threshold, this would correspond to a winding degree of about 9.

\subsection{Effects of rotation}

\begin{figure}[h]
    \centering

  \includegraphics[width=4.cm]{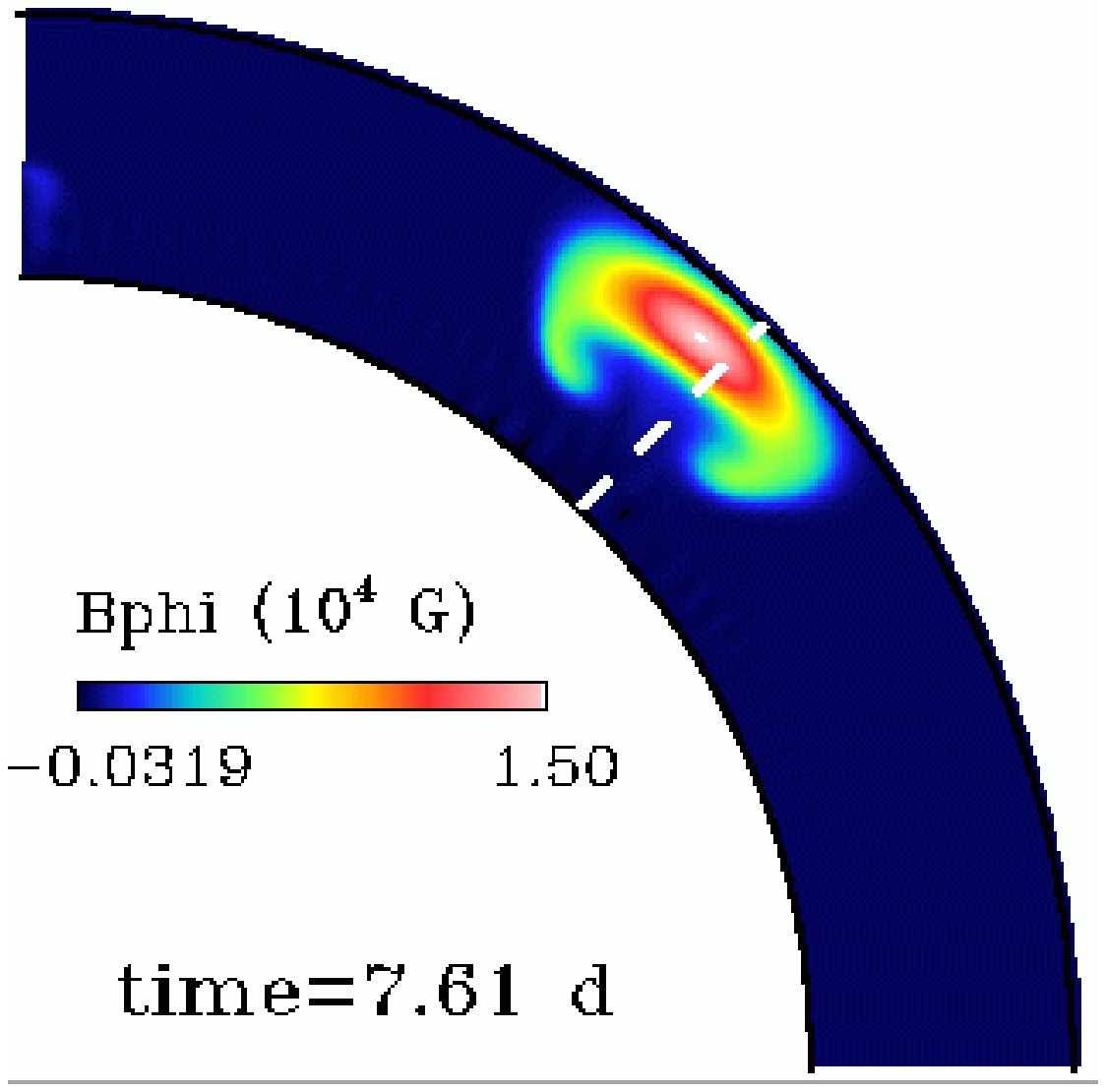}
  \includegraphics[width=4.cm]{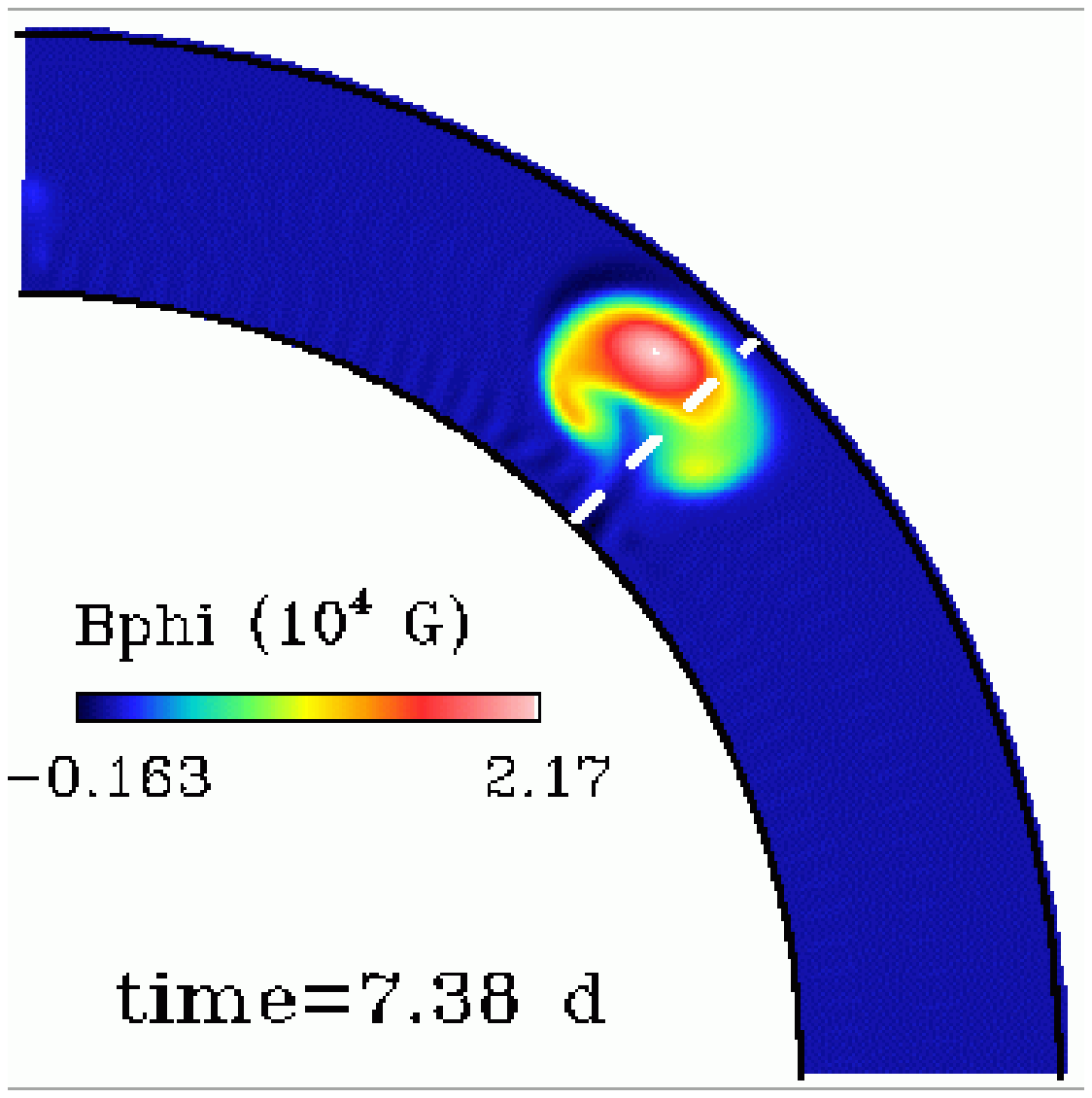}
       \caption{Cut at a constant longitude of $B_{\phi}$ after about 7 days of evolution of a flux tube initially located at the base of the CZ at $\theta=45\degr$ in the case of a non-rotating shell (left panel) and in a rotating case (right panel). The dashed lines represent the purely radial trajectory in each case.}
        \label{figure_rot1}
    \end{figure}

  To investigate the influence of rotation in our simulations, we compute a case where we remove the reference frame rotation. The initial latitude of the magnetic torus is still 45\degr and the tube is still introduced at the base of the convection zone, i.e. at about 0.75$R_\odot$. 
 Figure $\ref{figure_rot1}$ shows the behavior of the tube in the non-rotating case (left panels) and in the rotating case after about 7 days of evolution. 

 The most striking point is that in the non-rotating case, the tube remains much more symmetric with respect to the trajectory of the apex in a constant longitude plane. On the contrary, in the rotating case, we clearly see the distortion of the tube caused by the Coriolis force, the flux tends to be concentrated in the lobe which is closer to the rotation axis. 
    
    Figure \ref{figure_rot1} also shows the poleward slip of the flux tube characterized by the deviation to the purely radial trajectory represented by the dashed line. In the non-rotating case, the latitudinal component of the magnetic curvature acts to drag the tube poleward as it cannot be compensated by any equatorward force.  The deviation  angle to the radial direction is determined by comparing the latitude of the maximum of $B_{\phi}$ to its initial latitude. It is equal to 1.7$\degr$ in this case.
    In the rotating case, a retrograde zonal flow is created inside the tube which induces a Coriolis force directed towards the Sun's rotation axis which acts to deflect the trajectory of the tube poleward. Thus, we note that the deviation to the radial trajectory in this case is even more pronounced, reaching the value of about 3.8$ \degr $.
    Another striking difference between the two configurations resides in the rise velocity. Indeed, looking at Fig. $\ref{figure_rot1}$, we see that at  about the same instant, the tube in the rotating case has not risen as high in the CZ as the tube in the non rotating case. 
    
      \begin{figure}[h]
    \centering
  \includegraphics[width=8 cm]{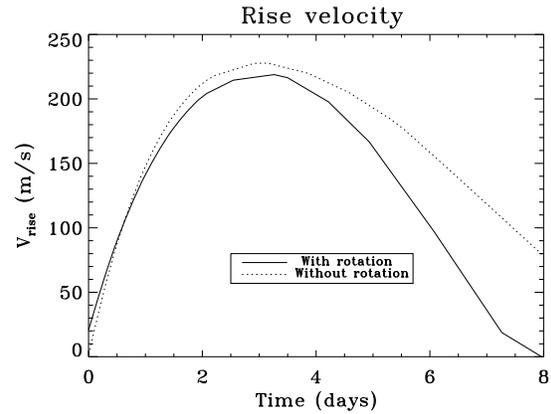}
       \caption{Rise velocity of the flux tube in the rotating and non-rotating cases.}
        \label{figure_velo}
    \end{figure}
    
       By following the position of the maximum of $B_\phi$ with respect to time, we are able to determine the profile of the rise velocity of the tube. Figure \ref{figure_velo} shows the evolution of the rise velocity of the flux tube in both cases.
    We note that the accelerating phase is similar, the maximum speed is about the same (about $220\,\rm m.s^{-1}$) but the decelerating phase is significantly modified in the rotating case. Indeed, the radial component of the centrifugal force decreases the tube velocity so that after 6 days of evolution, the rise velocity of the tube in the non-rotating case is about 1.5  times that of the tube in the rotating case. This can be explained by looking at the equation of evolution of the radial velocity in the simplified model of a thin axisymmetric flux tube in spherical geometry (Moreno-Insertis et al. 1992) which reads (assuming that  the rotation rate $\Omega$ of the tube is equal to the rotation rate of the surrounding medium and neglecting the advection terms):
    
    \begin{equation}
  \centering
  \frac{ \partial {v_r}}{\partial t}=-\frac{\Delta \rho}{\rho}(g-r\sin^2\theta\Omega^2)-\frac{B_{\phi}^2}{4\pi r \rho}+2\sin\theta v_{\phi}\Omega
  \label{eq_vr}
   \end{equation}
     
    The first term on the r.h.s of the equation is the buoyancy term which is modified by an extra term coming from the rotation and which has the effect of limiting the efficiency of buoyancy, thus resulting in a slower emergence in the rotating case.

\subsection{Influence of the initial latitude}

\begin{figure}[h]
    \centering
  \includegraphics[width=4.1cm]{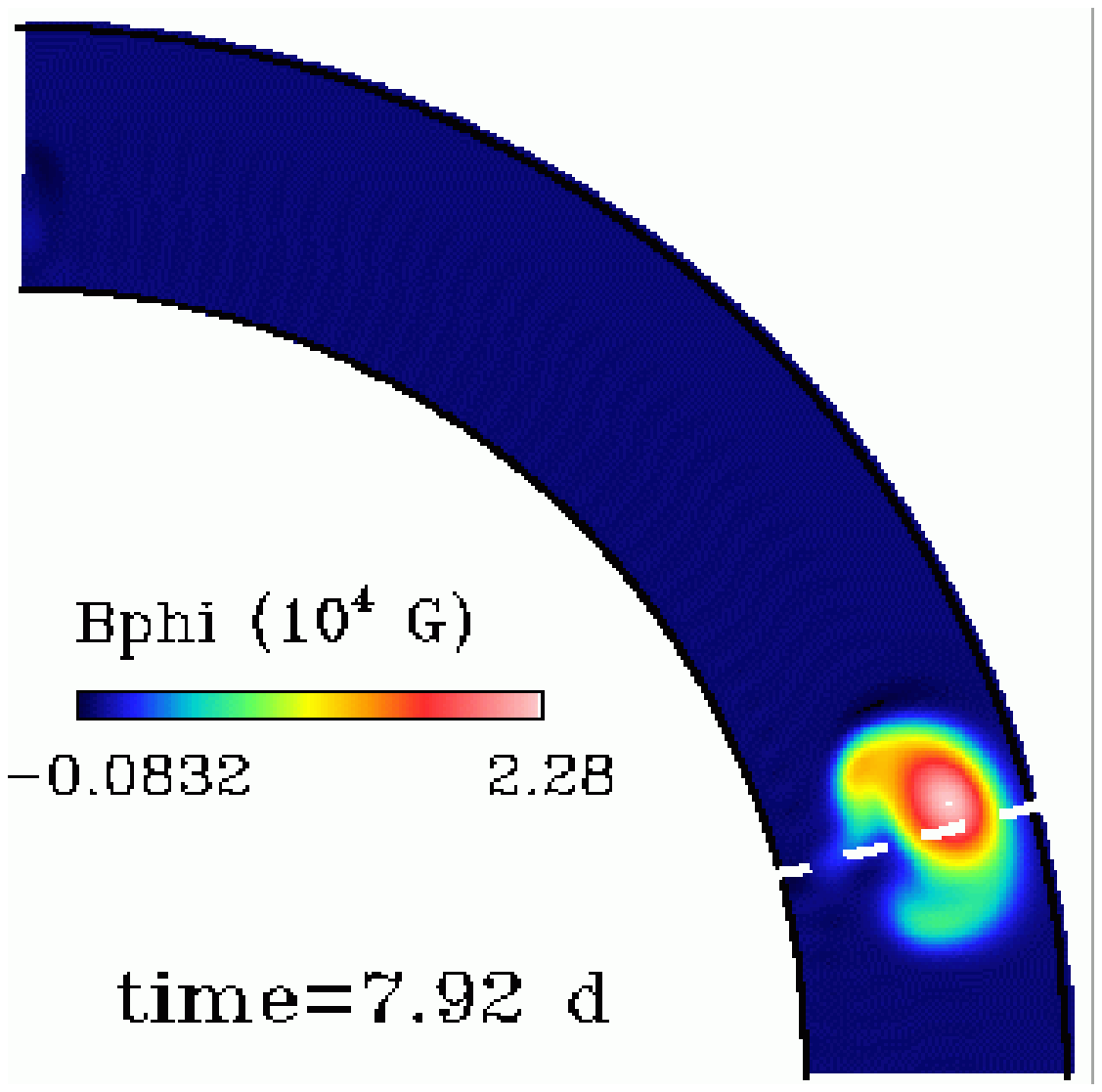}
  \includegraphics[width=4.1cm]{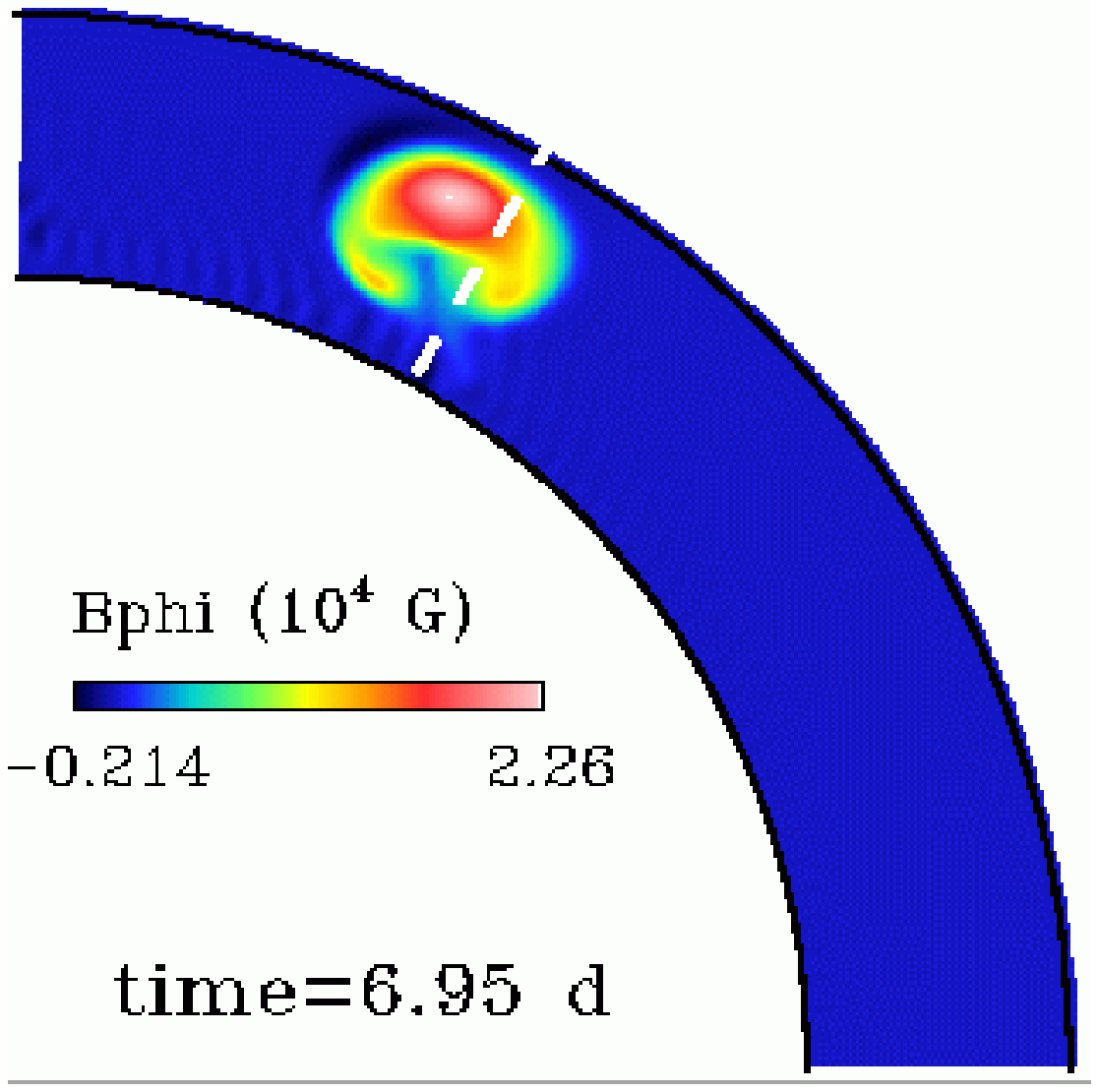}
       \caption{Same format as Fig. \ref{figure_rot1} for a flux tube initially located at the base of the CZ at a latitude of $15\degr$ and $60\degr$, in a rotating shell.}
        \label{figure_lat}
    \end{figure}

  Since our computations are made in a spherical shell, we are able to check the influence of spherical geometry on the dynamical evolution of the tube.  We can test for instance the influence of the initial latitude of the tube on its emergence. Figure \ref{figure_lat} shows the topology of $B_{\phi}$ in a constant longitude plane for flux tubes initially located at different latitudes.     
   The snapshots are chosen such that the tubes have reached the same height in the CZ. We note that the tube at the latitude of $15\degr$ is delayed by almost one day in comparison to the tube at the latitude of $60\degr$, we thus conclude that tubes located at higher latitudes, in a rotating background, are faster. In the non-rotating case, the rise velocity is the same for both tubes. It is again the modified buoyancy term of Eq.\ref{eq_vr} that is responsible for the lower rise velocity of the tube at low latitudes, since an increase in $\theta$ causes an increase of $r \sin^2\theta\,\,\Omega^2$ and thus a decrease of the efficiency of buoyancy to drive the tube upward. 
 
   Figure \ref{figure_lat} indicates that the initial latitude of the flux tube has also an influence on the deviation to the purely radial trajectory due to the poleward slip we mentioned before. In the case of a high latitude tube, the deviation reaches the value of 3.1$\degr$ whereas in the low latitude case, the poleward slip is small, reaching only the value of 1.5$\degr$. This result is also true for tubes introduced in a non-rotating shell and can be understood by looking at the very simplified model of an axisymmetric thin flux tube rising in a non-rotating shell. As Moreno-Insertis et al. (1992) indicate, we can understand the poleward drift in writing the equation for the $\theta$-component of the velocity in the non-rotating case, neglecting the advection terms:
   
  \begin{equation}
  \centering
  \frac{\partial v_{\theta}}{\partial t}=-\frac{B_{\phi}^2}{4\pi r \rho} \cot\theta
   \end{equation}
     
     This equation indicates that the acceleration in the $\theta$-direction is proportional to $\cot\theta$ which is a decreasing function of $\theta$ between $0$ and $\pi/2$. As $\theta$ is here the colatitude, the acceleration at higher latitudes is thus more rapidly active than at low latitudes and as a consequence, the poleward drift is much more visible for a flux tube originally located at a latitude of $60\degr$.

\section{Conclusion}

We have carried out the first study of the 3-D full MHD evolution in a spherical shell of a buoyant magnetic flux rope in an adiabatically stratified convective zone, with special emphasis on the effects of rotation and of the initial latitude of the tube.  We confirm several results obtained in different configurations such as in cartesian coordinates (e.g. Emonet \& Moreno-Insertis 1998) or within the thin flux tube approximation (e.g. Caligari, Moreno-Insertis \& Sch{\"u}ssler 1995). To be able to rise cohesively up to the solar surface, we confirm that a sufficient amount of twist of the field lines is needed. Moreover, we see in our simulations that twisted flux tubes can be strongly influenced by a background rotation which has two major effects: decreasing the rise velocity through the spherical shell and distorting the flux tube such that more flux gets concentrated closer to the rotation axis. In all these simulations, a poleward drift of the tube is observed due to the non-compensated curvature force linked to the magnetic tension existing in the tube.  Nevertheless, the spherical nature of these calculations enables to show that this poleward drift is minimized in the case where the tubes originate at low latitudes. This leads to consider that these low latitudes flux tubes could emerge at the surface in the observed solar activity belt. In this present work where flux tubes evolve in an isentropic background, 3-D effects do not play any role as our tubes remain axisymmteric during their rise. On the contrary, interactions with fully developed convection will lead to a modulation of the magnetic field in the azimuthal direction and thus to a creation of particularly active longitudes at the solar surface. The model has thus now to be improved especially by computing a fully developed convective case which will enable us to study the non-linear interactions of these flux tubes with self-consistently generated differential rotation, meridional flows and convective plumes. Such efforts are under way (Jouve \& Brun 2006, 2008).

\acknowledgements
We wish to thank the organizers of the Flux Emergence Workshop held in St Andrews in June 2007 and the 5th Potsdam Thinkshop for their invitations and for the very helpful discussions we had during the meetings.


\vspace{-0.3cm}

\begin{thebibliography}{}
\bibitem{}Brun, A.S., Miesch, S.M. \& Toomre, J. : 2004, ApJ, 614, 1073
\bibitem{}Caligari, P., Moreno-Insertis, F. \& Sch{\"u}ssler, M. : 1995, ApJ, 441, 886
\bibitem{}Cattaneo, F., Brummell, N. H. \& Cline, K. S. : 2006, MNRAS, 365, 727 
\bibitem{}Choudhuri, A. \& Gilman, P. : 1987, ApJ, 316, 788
\bibitem{}Emonet, T. \& Moreno-Insertis, F. : 1998, ApJ, 492, 804
\bibitem{}Fan, Y. : 2004, Living Rev. Solar Phys.
\bibitem{}Jouve, L., \& Brun, A.S. : 2006, SF2A-2006, 473
\bibitem{}Jouve, L., \& Brun, A.S. : 2008, in prep.
\bibitem{}Moreno-Insertis, F., Sch{\"u}ssler, M. \& Ferriz Mas, A. : 1992, A\&A, 264, 686
\bibitem{}Spruit, H.C. : 1981, A\&A, 98, 155
\bibitem{}Spruit, H.C., \& van Ballegooijen, A.A. : 1982, A\&A, 106, 58

\end{thebibliography}
\end{document}